\documentclass[twocolumn,showpacs,aps,prl,superscriptaddress,letterpaper]{revtex4}

\usepackage{graphicx}
\usepackage{dcolumn}
\usepackage{amsmath}
\usepackage{epsfig}
\usepackage{multirow}

\long\def\inst#1{\par\nobreak\kern 4pt\nobreak
    {\itshape #1}\par\vskip 10pt plus 3pt minus 3pt}
\RequirePackage{xspace}
\usepackage{relsize}

\newcommand{\BABARPubYear}     {09}
\newcommand{\BABARPubNumber}  {032}
\newcommand{\SLACPubNumber} {13898}

\input babarsym
\input hoobersym

\def\n1Spipi     {\ensuremath{}\xspace}

\def\beq{\begin{equation}}
\def\eeq{\end{equation}}
\def\bea{\begin{eqnarray}}
\def\eea{\end{eqnarray}}
\def\bq{\begin{quote}}
\def\eq{\end{quote}}
\def\bi{\begin{itemize}}
\def\ei{\end{itemize}}
\def\bc{\begin{center}}
\def\ec{\end{center}}

\begin{document}

\preprint{\babar-PUB-\BABARPubYear/\BABARPubNumber} 
\preprint{SLAC-PUB-\SLACPubNumber} 

\begin{flushleft}
\babar-PUB-\BABARPubYear/\BABARPubNumber\\
SLAC-PUB-\SLACPubNumber\\
\end{flushleft}


\title{ \large \bfseries \boldmath  Search for Charged Lepton Flavor Violation in Narrow Upsilon Decays}

%
\author{J.~P.~Lees}
\author{V.~Poireau}
\author{E.~Prencipe}
\author{V.~Tisserand}
\affiliation{Laboratoire d'Annecy-le-Vieux de Physique des Particules (LAPP), Universit\'e de Savoie, CNRS/IN2P3,  F-74941 Annecy-Le-Vieux, France}
\author{J.~Garra~Tico}
\author{E.~Grauges}
\affiliation{Universitat de Barcelona, Facultat de Fisica, Departament ECM, E-08028 Barcelona, Spain }
\author{M.~Martinelli$^{ab}$}
\author{A.~Palano$^{ab}$ }
\author{M.~Pappagallo$^{ab}$ }
\affiliation{INFN Sezione di Bari$^{a}$; Dipartimento di Fisica, Universit\`a di Bari$^{b}$, I-70126 Bari, Italy }
\author{G.~Eigen}
\author{B.~Stugu}
\author{L.~Sun}
\affiliation{University of Bergen, Institute of Physics, N-5007 Bergen, Norway }
% ...
\collaboration{The \babar\ Collaboration}
\noaffiliation

\begin{abstract}
Charged  lepton flavor  violating processes  are  unobservable in  the
standard  model,   but  they  are   predicted  to be enhanced   in  several
extensions to the standard model,  including  supersymmetry  and
models  with leptoquarks or  compositeness.  We  present a  search for
such processes in a sample of $99\times10^6$ $\Upsilon(2S)$ decays and
$117\times10^6$  $\Upsilon(3S)$  decays  collected  with  the  \babar\
detector.  We place upper  limits on the branching fractions \bfnsetau
and \bfnsmutau $(n=2,3)$ at the  $10^{-6}$ level and use these results
to place lower  limits of order 1~TeV on the mass scale of charged 
lepton flavor violating effective operators.
\end{abstract}

\pacs{ 
11.30.Hv, 
13.20.Gd, 
14.40.Nd  
}

\maketitle


In  the original  formulation  of  the standard  model  (SM) in  which
neutrinos  are massless,  lepton flavor  is an  accidentally conserved
quantum number.   The extension of  the SM to include  neutrino masses
introduces  oscillations between the  neutrino flavors,  which violate
this conservation law.  However, SM processes involving charged lepton
flavor   violation  (CLFV)  remain   unobservable  because   they  are
suppressed by the quantity $(\Delta m_{\nu}^2/M_W^2)^2 \lsim 10^{-48}$
\cite{ref:feinberg,ref:bilenky,ref:mnu}.   Here $\Delta  m_{\nu}^2$ is
the difference  between the squared  masses of neutrinos  of different
flavor and  $M_W$ is the charged  weak vector boson  mass.  Hence CLFV
represents  an  unambiguous signature  of  new  physics
(NP)~\cite{ref:lfvA,ref:lfvB,ref:pati,ref:georgi}.  Many extensions to
the  SM,  including  supersymmetry  and  models  with  leptoquarks  or
compositeness, predict an enhancement in the rates for these processes
at  levels   close  to  experimental  sensitivity.    There  have  been
considerable efforts in  searches for CLFV in decays  of particles such
as $\mu$  and $\tau$ leptons and $B$  and $K$ mesons, but  CLFV in the
$\Upsilon$  sector remains relatively  unexplored~\cite{ref:cleo}.  By
using  unitarity  considerations,   limits  on  CLFV  $\tau$ branching
fractions~\cite{ref:bellelfv} have been used  to place indirect limits on
CLFV  $\Upsilon$  branching  fractions at  the  $\mathcal{O}(10^{-3})$
level~\cite{ref:nussinov}.  In  this letter  we describe a  search for
CLFV $\Upsilon$ decays,  which is a thousand times  more sensitive than
these indirect limits, using  data collected with the \babar\ detector
at  the \pep2\ $B$  factory at  SLAC National  Accelerator Laboratory.
Since these decays are in  general mediated by new
particles produced  off-shell in loops, their  measurement probes mass
scales up to the TeV  scale, far exceeding the $e^+e^-$ center-of-mass
(CM)                          collision                         energy
$\sqrt{s}=M_{\Upsilon(nS)}=\mathcal{O}(10~\mathrm{GeV})$~\cite{ref:barbieri}.
Therefore, this analysis provides a NP probe which is complementary to
direct searches  ongoing at  the Tevatron and  to be performed  at the
Large Hadron Collider.

Assuming  that  the partial  widths  for  CLFV  $\Upsilon$ decays  are
comparable  at  the \Y2S,  \Y3S  and  \Y4S  resonances, the  branching
fractions  for rare  decays  of the  narrow $\Upsilon(nS)$  resonances
(henceforth   $n\equiv    2,3$)   are   enhanced    by   approximately
$\Gamma_{\Upsilon(4S)}         /        \Gamma_{\Upsilon(nS)}        =
\mathcal{O}(10^3)$~\cite{ref:pdg}  with   respect  to  those   of  the
$\Upsilon(4S)$.    We  search   for  the   CLFV  decays   \nsetau  and
\nsmutau~\cite{ref:thesis},  while the  decay $\Upsilon(nS)\rightarrow
e^{\pm}\mu^{\mp}$  is constrained  by unitarity  considerations  to be
less  than $\mathcal{O}(10^{-8})$~\cite{ref:nussinov}.   No  signal is
expected  in data  collected  at  the \Y4S  since  the CLFV  branching
fractions are strongly suppressed, or  in data collected away from the
$\Upsilon$ resonances,  since this  data contains very  few $\Upsilon$
decays.  We  search for CLFV  in a sample  of $(98.6\pm0.9)\times10^6$
$\Upsilon(2S)$  decays  and  $(116.7\pm1.2)\times10^6$  $\Upsilon(3S)$
decays  corresponding to  integrated luminosities  of  13.6~\invfb and
26.8~\invfb, respectively.  Data collected at the $\Upsilon(4S)$ after
the  upgrade  of  the  muon  detector system  (77.7~\invfb)  and  data
collected   30~MeV  below   the   $\Upsilon(2S)$  and   $\Upsilon(3S)$
resonances (off-peak data  corresponding to 2.6~\invfb and 1.3~\invfb,
respectively) constitute control samples that are used to validate the
fit  procedure.  An additional  data control  sample collected  at the
\Y3S  resonance  (1.2~\invfb)  is  used  in  a  preliminary  unblinded
analysis to  validate the analysis  procedure and to  ensure agreement
between data  and events simulated using Monte  Carlo (MC) techniques.
Simulated   background   processes   consisting   of   continuum   QED
events~\cite{ref:bhwide,ref:kk2f}  and generic  $\Upsilon(nS)$ decays,
as     well     as     signal     \nsltau     ($\ell\equiv     e,\mu$)
decays~\cite{ref:evtgen},  are produced and  analyzed to  optimize the
fit procedure.  The  \textsc{Geant4}~\cite{ref:geant} software is used
to   simulate   the   interactions   of   particles   traversing   the
\babar\     detector,     which     is     described     in     detail
elsewhere~\cite{ref:babar}.


The   signature   for  \nsltau   events   consists   of  exactly   two
oppositely charged particles: a primary lepton, an electron (muon) for
the \nsetau (\nsmutau) search, with  momentum close to the beam energy
$E_B=\sqrt{s}/2$, and a secondary  charged lepton or charged pion from
the  $\tau$ decay.   Here  and  in the  following  all quantities  are
defined in  the CM  frame unless otherwise  specified.  If  the $\tau$
decays  to leptons, we  require that  the primary  and $\tau$-daughter
leptons are of different flavor.   If the $\tau$ decays to hadrons, we
require one  or two additional  neutral pions from this  decay.  These
requirements  on  the  identified  particle  types  are  necessary  to
suppress  Bhabha  and $\mu$-pair  backgrounds.   Thus  we define  four
signal  channels, consisting  of  leptonic and  hadronic $\tau$  decay
modes for the \nsetau and  \nsmutau searches, hereafter referred to as
the leptonic  and hadronic $e\tau$  and $\mu\tau$ channels.   The main
source of background to  our events comes from $\tau$-pair production,
for which  the final state particles  are the same as  for the signal.
There is a background contribution to the $e\tau$ channels from Bhabha
events  in which one  of the  electrons is  misidentified, and  to the
$\mu\tau$ channels from $\mu$-pair events in which one of the muons is
misidentified or  decays in flight, or  an electron is  generated in a
material interaction.   An additional background  consisting of events
with  multiple pions  and possible  additional  photons (`$\pi$-hadron
background'), in which a charged pion is misidentified as a lepton and
the  remaining particles pass  the selection  criteria for  the $\tau$
decay  products, contributes  to  the hadronic  $e\tau$ and  $\mu\tau$
channels.

In order to  reduce background, we first apply  requirements common to
all the decay modes and then a channel specific selection.  All events
are  required to  have exactly  two  tracks of  opposite charge,  both
consistent  with originating  from the  primary interaction  point and
with opening angle greater  than 90$^{\circ}$.  To suppress Bhabha and
$\mu$-pair backgrounds, we require that $M_{vis}/\sqrt{s}<0.95$, where
$M_{vis}$ is the invariant mass of the sum of the 4-vectors of the two
charged  particles and  of all  photon  candidates in  the event.   To
ensure that the  missing momentum is not pointing  toward the holes in
the    detector     near    the    beamline,     we    require    that
$\cos(\theta^{lab}_{miss})<0.9$  and  $\cos(\theta^{CM}_{miss})>-0.9$,
where $\theta^{lab}_{miss}$ ($\theta^{CM}_{miss}$)  is the polar angle
of the missing momentum in  the lab (CM) frame. To suppress two-photon
processes,                we                require               that
$(\boldsymbol{p_1}+\boldsymbol{p_2})_{\perp}/(\sqrt{s}-|\boldsymbol{p_1}|-|\boldsymbol{p_2}|)>0.2$,
where $\boldsymbol{p_1}$ and $\boldsymbol{p_2}$ are the momenta of the
two charged  particles and $\perp$ indicates  the transverse component
with respect to the beam axis.

Particle  identification is  performed using  a  multivariate analysis
\cite{ref:narsky}  which uses  measurements from  all of  the detector
subsystems.   An electron selector  and a muon  veto, combined  with the
requirement that  the particle falls within the  angular acceptance of
the electromagnetic calorimeter (EMC), are used to identify electrons.
A muon selector and an electron veto  are used to identify muons, while a
charged  pion  selector, an electron  veto  and a muon  veto are  used  to
identify charged  pions.  The particle  misidentification efficiencies
are         $\mathcal{O}(10^{-6})$        $(\mu\rightarrow        e)$,
$\mathcal{O}(10^{-5})$ $(e\rightarrow \mu)$ and $\mathcal{O}(10^{-1})$
$(\ell\rightarrow  \pi)$,  where  $x\rightarrow  y$ indicates  that  a
particle of type $x$ is misidentified as a particle of type $y$. A photon
candidate must  deposit at least 50~MeV  in the EMC and  have a shower
profile consistent with that  expected from an electromagnetic shower.
All pairs of photons with an invariant mass between 0.11~GeV and 0.16~GeV are
selected as neutral pion candidates.

The channel specific selection classifies  events into one of the four
signal channels. The momentum of  the primary lepton normalized to the
beam        energy         is        required        to        satisfy
$x\equiv|\boldsymbol{p_1}|/E_B>0.75$.   For the  hadronic $\tau$-decay
channels, the momentum of the $\tau$-daughter charged pion is required
to satisfy $|\boldsymbol{p_2}|/E_B<0.8$.  Since these $\tau$ decays to
hadronic  final  states  are   dominated  by  the  decays  $\tau^{\pm}
\rightarrow \rho^{\pm} \nu_\tau$ and $\tau^{\pm} \rightarrow a_1^{\pm}
\nu_\tau$,     the    masses     of    the     $\pi^{\pm}\pi^0$    and
$\pi^{\pm}\pi^0\pi^0$ systems  are required to be  consistent with the
masses  $m_{\rho}=0.77~\mathrm{GeV}$  and $m_{a_1}=1.26~\mathrm{GeV}$,
respectively,  where  the  requirement  on  the  $\pi^{\pm}\pi^0\pi^0$
system mass  is included only  if there are  two neutral pions  in the
event.  In  order to  suppress Bhabha events  in which an  electron is
misidentified  as  a  muon,  for  the  leptonic  $e\tau$  channel  the
$\tau$-daughter  muon is required  to penetrate  deeply into  the muon
detector.  In order to suppress  $\mu$-pair events in which the tracks
are back-to-back,  for the leptonic  $\mu\tau$ channel the  tracks are
required  to satisfy  $\Delta\phi<172^{\circ}$, where  $\Delta\phi$ is
the difference  between the  track azimuthal angles.   After including
all  selection requirements,  typical  signal efficiencies  determined
from  MC are  $(4-6)\%$~\cite{ref:suppl}, including  the  $\tau$ decay
branching  fractions.   The  typical  number  of  events  passing  the
selection   criteria  is   $(10-15)\times10^3$  for   \Y2S   data  and
$(20-30)\times10^3$ for  \Y3S data,  depending on the  signal channel.
These yields are consistent with background expectations from MC simulations.


After  selection  an  unbinned  extended maximum  likelihood  fit  is
performed to  the distribution  of the discriminant  variable $x$.
The signal  peaks at  $x\approx0.97$, while  the $\tau$-pair
background  $x$   distribution  is  smooth  and   approaches  zero  as
$x\rightarrow  x_{MAX}$, where  $x_{MAX}\approx0.97$ is  the effective
kinematic  endpoint  for the  lepton  momentum  in  the decay  $\tau^{\pm}
\rightarrow  \ell^{\pm}  \bar{\nu_{\ell}}\nu_{\tau}$,  boosted  into  the
$\Upsilon(nS)$   rest-frame.    The    $x$   distributions   for   the
Bhabha/$\mu$-pair  backgrounds have  a peaking  component  near $x=1$,
about    (2.5-3)$\sigma_x$    above    the    signal    peak,    where
$\sigma_x\approx0.01$  denotes the  detector $x$ resolution. The
$x$ distribution  for the $\pi$-hadron background is  smooth and falls
off  sharply near $x=x_{MAX}$.   Probability density  functions (PDFs)
for   signal,    $\tau$-pair,   Bhabha/$\mu$-pair   and   $\pi$-hadron
backgrounds are determined as discussed below, and a PDF consisting of
the sum of these components weighted  by their yields is fitted to the
data  for each  signal  channel,  with the  yields  of the  components
allowed to vary in the fit.

The PDFs  for signal  and Bhabha/$\mu$-pair backgrounds  are extracted
from  fits to  the  $x$ distributions  of  MC events.   The signal  is
modeled  by  a modified  Gaussian  with  low-  and high-energy  tails,
hereafter referred to as a double Crystal Ball~\cite{ref:cb} function,
which peaks near $x=0.97$.  The Bhabha and $\mu$-pair backgrounds have
a threshold  component truncating near  $x=1$, which is modeled  by an
ARGUS  distribution~\cite{ref:argus},  and  a peaking  component  near
$x=1$,  which  is modeled  by  a  Gaussian function.  The $\pi$-hadron  PDF  is
determined from  data by modifying  the selection to require  that the
primary lepton is instead identified as a charged pion.  The resulting
binned $x$ distribution  is scaled by the probability  for pions to be
misreconstructed as charged  leptons, as measured in data,  to yield a
binned  PDF  for  the  $\pi$-hadron  background.  The  yield  of  this
component is fixed in the maximum likelihood fit and an uncertainty of
10\%  is  assessed.  The  $\tau$-pair  background  is  modeled by  the
convolution  of  a  polynomial,  which vanishes  above  the  kinematic
endpoint  $x_{MAX}$, and  a  detector resolution  function.
The detector resolution  function is modeled by a  double Crystal Ball
function whose  shape is extracted from $\tau$-pair  MC events.  Since
the  signal peaks in  the region  near the  kinematic endpoint  of the
$\tau$-pair  background  $x$ distribution,  the  signal yield  depends
strongly on  $x_{MAX}$, which must  therefore be extracted  from data.
The  value   of  this  parameter   is  extracted  from  fits   to  the
$\Upsilon(4S)$ data  control sample  and corrected for  differences in
the decay kinematics at  the \Y4S vs.  $\Upsilon(nS)$ resonances.  The
polynomial shape  parameters, which  are not strongly  correlated with
the signal  yield, are allowed to  vary in the  fits to $\Upsilon(nS)$
data.

\begin{figure}[!t]
\begin{center}
\epsfig{file=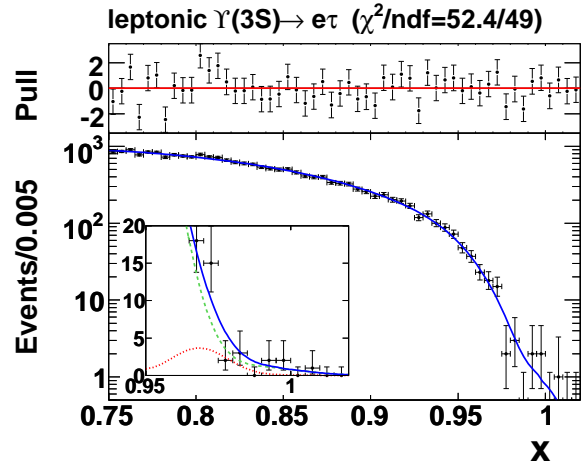, width=0.48\textwidth}
\end{center}
\vspace{-2\baselineskip}
\caption{Maximum  likelihood  fit  results  for the  leptonic  $e\tau$
  channel in \Y3S data. The  red dotted line represents the signal PDF,
  the green dashed line represents the sum of  all background PDFs and
  the solid blue line represents the sum of these components. The inset
  shows a close-up of the region $0.95<x<1.02$. 
  The top plot shows the normalized residuals
  $(\mathrm{data} - \mathrm{fit})/\sigma_{\mathrm{data}}$ (pull).}
\vspace{-1\baselineskip}
\label{fitresults}
\end{figure}

To validate the fit procedure, we perform fits to data control samples
in  order  to verify  that  signal  yields  consistent with  zero  are
obtained.  The  $\Upsilon(4S)$ data is  divided into samples  that are
chosen to be comparable in size to the \Y2S and \Y3S data samples. The
off-peak data  and the 1.2~\invfb  of \Y3S data  constitute additional
data control  samples.  Results consistent with zero  signal yield are
obtained for  all signal channels  in these data control  samples.  

The branching fraction $\mathcal{B}$  is calculated from the extracted
signal          yield          $N_{SIG}$         according          to
$\mathcal{B}=N_{SIG}/(\epsilon_{SIG}\times  N_{\Upsilon(nS)})$,  where
$\epsilon_{SIG}$    is   the    signal   selection    efficiency   and
$N_{\Upsilon(nS)}$ is  the number of  collected $\Upsilon(nS)$ decays.
The dominant systematic uncertainties  in the signal yields, which arise from
uncertainties in the  PDF shapes, are determined  by varying the
shape parameters  while taking into account the  correlations between them.
This uncertainty is  3-10 events depending on the  signal channel, and
the largest  contribution is due  to the uncertainty in  the kinematic
endpoint parameter $x_{MAX}$.  To assess the uncertainty in the signal
efficiency, we take the relative difference between the yields for data and
MC  events from  a portion  of the  sideband of  the  $x$ distribution
defined  by $0.8<x<0.9$,  which  is dominated  by $\tau$-pair  events.
This difference is due  to particle identification, tracking, trigger,
and  kinematic  selection   efficiency  uncertainties.   There  is  an
additional statistical  uncertainty in the signal  efficiency, as well
as  an uncertainty  arising from  the uncertainties in the  $\tau$ branching
fractions~\cite{ref:pdg}.   The total signal  efficiency uncertainties
are $(2-4)$\%,  depending on the  signal channel.  The  uncertainty on
the number of collected $\Upsilon(nS)$ decays is approximately 1\%.

To  assess the  possible bias  in the  fit procedure,  several hundred
simulated  experiments are  produced with  the generated  signal yield
fixed  to   the  larger  of  the   value  extracted  by   the  fit  to
$\Upsilon(nS)$ data, or zero.  The bias is consistent with zero within
the uncertainty  of 0.2-0.7 events,  depending on the  signal channel.
There is also an uncertainty resulting from a correction in the signal
yield  which is  performed  to compensate  for  primary leptons  whose
momentum is poorly measured. These particles populate a broad momentum
range  and some  fall in  the signal  region defined  as  the interval
within $\pm1.5\sigma$ of the signal  peak.  The number of these events
is estimated using $\tau$-pair  MC simulation and corrected using data
and  MC   Bhabha  and   $\mu$-pair  control  samples.    The  expected
contributions are  subtracted from the signal yields  extracted by the
fit and an uncertainty of 100\% times the correction is assessed.  The
corrections are  approximately 3 events (5 events)  for the \twosmutau
(\threesmutau)  channels  and  less  than  1  event  for  the  \nsetau
channels.


The maximum likelihood fit results  for a sample channel are displayed
in  Fig.~\ref{fitresults} and  the fit  results for  all  channels are
available   in~\cite{ref:suppl}.   After  including   statistical  and
systematic uncertainties, the extracted signal yields for all channels
are consistent  with zero within $\pm1.8\sigma$.  We  conclude that no
statistically  significant  signal  is  observed  and  determine  90\%
confidence level (CL) upper limits (UL) using a Bayesian technique, in
which  the prior likelihood  is uniform  in $\mathcal{B}$  and assumes
that    $\mathcal{B}>0$.    The    resulting   ULs,    summarized   in
Table~\ref{finalresultstable},    are    $\mathcal{O}(10^{-6})$    and
represent   the   first   constraints   on   $\mathcal{B}(\Upsilon(nS)
\rightarrow   e^{\pm}\tau^{\mp})$.     These   results   improve   the
sensitivity by factors  of 3.7 and 5.5, respectively,  with respect to
the    previous   ULs    on    $\mathcal{B}(\Upsilon(2S)   \rightarrow
\mu^{\pm}\tau^{\mp})$    and   $\mathcal{B}(\Upsilon(3S)   \rightarrow
\mu^{\pm}\tau^{\mp})$~\cite{ref:cleo}.

\begin{table}
\caption{Branching fractions and 90\% CL ULs
for signal decays. The first error is statistical and the second is systematic.}
\begin{center}
\begin{tabular}{lcc}
\hline  & \raisebox{-1ex}{$\mathcal{B}$ $(10^{-6})$} & \raisebox{-1ex}{UL $(10^{-6})$} \\[2ex]  
\hline \raisebox{-1ex}{\bftwosetau} &  \raisebox{-1ex}{$0.6_{-1.4-0.6}^{+1.5+0.5}$} & \raisebox{-1ex}{$<3.2$} \\[2ex]  
\raisebox{-1ex}{\bftwosmutau}       &  \raisebox{-1ex}{$0.2_{-1.3-1.2}^{+1.5+1.0}$} & \raisebox{-1ex}{$<3.3$} \\[2ex]  
\raisebox{-1ex}{\bfthreesetau}      &  \raisebox{-1ex}{$1.8_{-1.4-0.7}^{+1.7+0.8}$} & \raisebox{-1ex}{$<4.2$} \\[2ex]
\raisebox{-1ex}{\bfthreesmutau}  &  \raisebox{-1ex}{$-0.8_{-1.5-1.3}^{+1.5+1.4}$} & \raisebox{-1ex}{$<3.1$} \\[2ex] 
\hline
\end{tabular}
\vspace{-2\baselineskip}
\label{finalresultstable}
\end{center}
\end{table}

\begin{figure}[!t]
\begin{center}
\epsfig{file=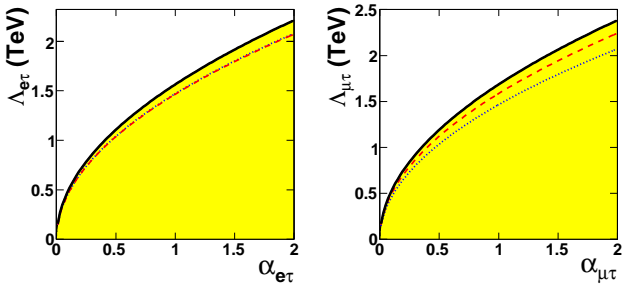, width=0.48\textwidth}
\end{center}
\vspace{-2\baselineskip}
\caption{Excluded regions of effective field theory  parameter  spaces  of  
mass  scale  $\Lambda_{\ell\tau}$
versus coupling constant  $\alpha_{\ell\tau}$. The  dotted blue  line is
derived from  \Y2S results only, the  dashed red line  is derived from
\Y3S results  only, and  the solid black  line indicates  the combined
results.  The yellow shaded  regions are  excluded at  90\% CL.}
\vspace{-1\baselineskip}
\label{exclusion}
\end{figure}

Our  results  can  be  used  to constrain  NP  using  effective  field
theory. The CLFV $\Upsilon(nS)$ decays may be parameterized as an effective
$b\bar{b}\ell^{\pm}\tau^{\mp}$  4-fermion interaction given by~\cite{ref:cleo}
\begin{equation}
\Delta\mathcal{L}=\frac{4\pi\alpha_{\ell\tau}}{\Lambda_{\ell\tau}^2}
(\bar{\ell}\Gamma_{\mu}\tau)(\bar{b}\gamma^{\mu}b),
\end{equation}
where $\Gamma_{\mu}$ is a vector or an axial current or their combination,
$\alpha_{\ell\tau}$ and $\Lambda_{\ell\tau}$ are the NP
coupling constant and mass scale, respectively.
This allows the following relation to be derived~\cite{ref:silagadze,ref:black}:
\begin{equation}
\frac                      {\alpha_{\ell\tau}^2}{\Lambda_{\ell\tau}^4}=
\frac{\mathcal{B}(\Upsilon(nS)    \rightarrow   \ell^{\pm}\tau^{\mp})}
     {\mathcal{B}(\Upsilon(nS)   \rightarrow   \ell^+\ell^-)}  \frac{2
       q_b^2 \alpha^2}{(M_{\Upsilon(nS)})^4}.  \nonumber
\label{lfvequation}
\end{equation}
Here    $q_b=-1/3$    is    the    charge   of    the    $b$    quark,
$\alpha\equiv\alpha(M_{\Upsilon(nS)})$ is  the fine structure constant
evaluated  at  the  $\Upsilon(nS)$  mass,  and we  take  the  dilepton
branching      fraction      $\mathcal{B}(\Upsilon(nS)     \rightarrow
\ell^+\ell^-)$ from  the average of the  $\Upsilon(nS)$ dielectron and
dimuon  branching fractions~\cite{ref:pdg}.   Using  these values  and
taking into  account the 7\%  uncertainty in $\mathcal{B}(\Upsilon(nS)
\rightarrow \ell^+\ell^-)$, we determine  the likelihood as a function
of the quantity $\alpha_{\ell\tau}^2/\Lambda_{\ell\tau}^4$ and extract
the 90\% CL UL using the  same Bayesian method as above.  We use these
results   to   exclude  regions   of   the  $\Lambda_{\ell\tau}$   vs.
$\alpha_{\ell\tau}$ parameter spaces as shown in Fig.~\ref{exclusion}.
Assuming $\alpha_{e\tau}=\alpha_{\mu\tau}=1$,  these results translate
to   the   90\%   CL   lower  limits   $\Lambda_{e\tau}>1.6$~TeV   and
$\Lambda_{\mu\tau}>1.7$~TeV on  the mass  scale of NP  contributing to
CLFV  $\Upsilon(nS)$ decays,  which  improve upon  the previous  lower
limit on $\Lambda_{\mu\tau}$~\cite{ref:cleo}.

We are grateful for the excellent luminosity and machine conditions
provided by our \pep2\ colleagues, 
and for the substantial dedicated effort from
the computing organizations that support \babar.
The collaborating institutions wish to thank 
SLAC for its support and kind hospitality. 
This work is supported by
DOE
and NSF (USA),
NSERC (Canada),
CEA and
CNRS-IN2P3
(France),
BMBF and DFG
(Germany),
INFN (Italy),
FOM (The Netherlands),
NFR (Norway),
MES (Russia),
MEC (Spain), and
STFC (United Kingdom). 
Individuals have received support from the
Marie Curie EIF (European Union) and
the A.~P.~Sloan Foundation.

\onecolumngrid
\newpage

\section{Appendix: EPAPS Material}

The following includes supplementary material for the Electronic
Physics Auxiliary Publication Service. 

\begin{table}[h]
\caption{Signal efficiencies $\epsilon_{SIG}$ and signal yields $N_{SIG}$ extracted by 
the maximum likelihood fit for the four
signal channels in \Y2S and \Y3S data. The first error  is statistical and the second is systematic.}
\begin{center}
\begin{tabular}{llcccc}
\hline
     &                       & leptonic $e\tau$ & hadronic $e\tau$ & leptonic $\mu\tau$ & hadronic $\mu\tau$ \\
\hline
\Y2S & $\epsilon_{SIG}~(\%)$ & $5.10\pm0.03\pm0.10$ & $5.53\pm0.03\pm0.16$ & $4.22\pm0.03\pm0.11$ & $6.08\pm0.03\pm0.12$  \\
     & $N_{SIG}$             & $-8\pm8\pm3$         & $19\pm12\pm4$        & $-11\pm8\pm5$        & $17\pm13\pm7$         \\
\hline            
\Y3S & $\epsilon_{SIG}~(\%)$ & $5.24\pm0.05\pm0.17$ & $5.56\pm0.05\pm0.12$ & $4.20\pm0.05\pm0.06$ & $6.07\pm0.06\pm0.10$  \\ 
     & $N_{SIG}$             & $24\pm13\pm4$        & $-6\pm13\pm6$        & $-11\pm10\pm9$       & $4\pm16\pm11$         \\    
\hline
\end{tabular}
\vspace{-2\baselineskip}
\label{yieldtable}
\end{center}
\end{table}

\begin{figure*}[!ht]
\begin{center}
\begin{tabular}{cc}
\epsfig{file=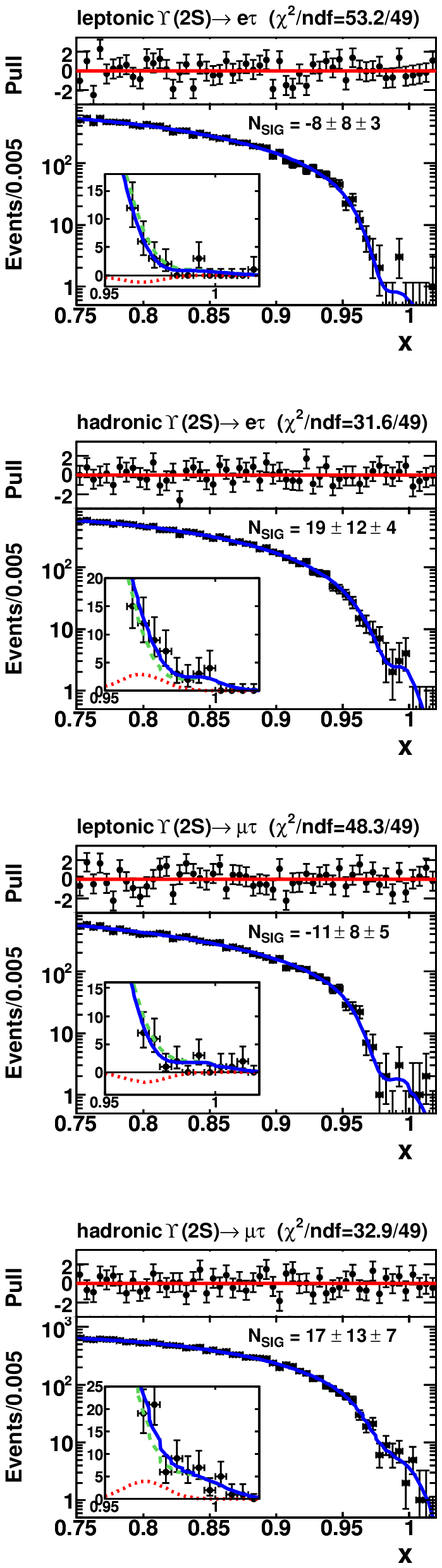, width=0.27\textwidth} &
\epsfig{file=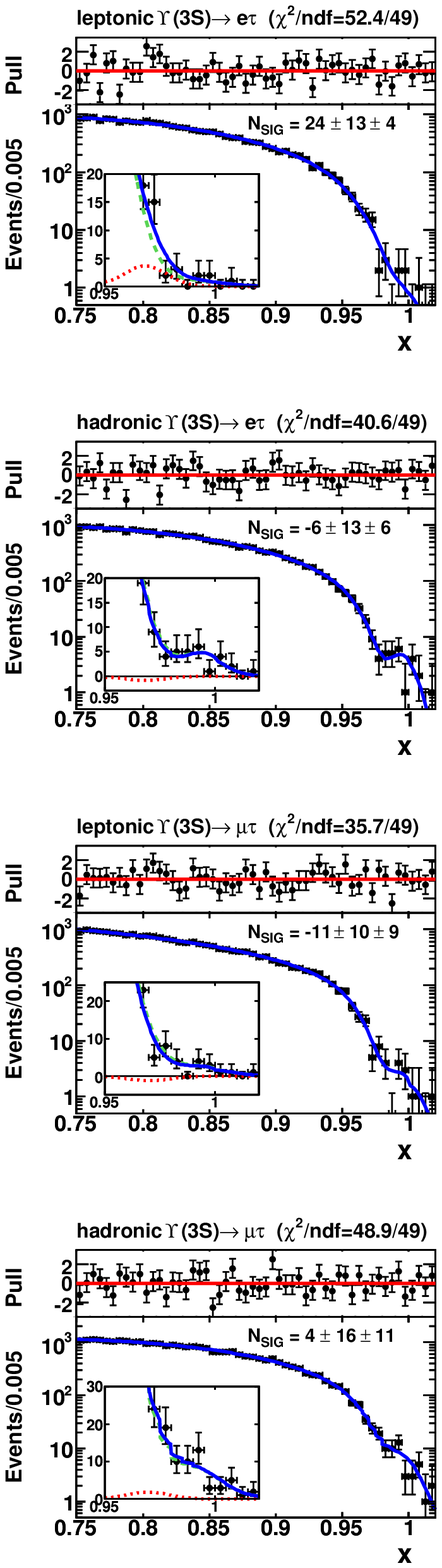, width=0.27\textwidth} \\
\end{tabular}
\end{center}
\vspace{-2\baselineskip}
\caption{Maximum likelihood  fit results for the  four signal channels
in  \Y2S data  (left)  and \Y3S  data  (right).  The  red dotted  line
indicates the signal  PDF, the green dashed line  indicates the sum of
all background PDFs and the solid blue line indicates the sum of these
components.   The    inset   shows   a   close-up    of   the   region
$0.95<x<1.02$. The pull denotes the difference in each bin between the
data  and total  PDF,  normalized to  the  statistical uncertainty  in
data.  The   corrected  signal  yield  $N_{SIG}$   is  displayed  with
statistical and systematic errors.}
\vspace{-1\baselineskip}
\label{allfitresults}
\end{figure*}


\begin{thebibliography}{99}

\bibitem{ref:feinberg}
G.\ Feinberg, Phys.\ Rev. \textbf{110}, 1482 (1958).

\bibitem{ref:bilenky}
S.\ M.\ Bilenky and B.\ Pontecorvo, Phys.\ Lett. \textbf{B61}, 248 (1976).

\bibitem{ref:mnu}
A.\ Strumia and F.\ Vissani, arXiv:hep-ph/0606054, (2007).

\bibitem{ref:lfvA}
J.\ R.\ Ellis {\em et al.}, Eur.\ Phys.\ J. \textbf{C14}, 319 (2000).

\bibitem{ref:lfvB}
J.\ R.\ Ellis, M.\ Raidal and T.\ Yanagida, Phys.\ Lett. \textbf{B581}, 9 (2004).

\bibitem{ref:pati}
J.\ C.\ Pati and A.\ Salam, Phys.\ Rev. D\ \textbf{10}, 275 (1974)
and {\em erratum} Phys.\ Rev. D\ \textbf{11}, 703 (1975).

\bibitem{ref:georgi}
H.\ Georgi and S.\ L.\ Glashow, Phys.\ Rev.\ Lett. \textbf{32}, 438 (1974).

\bibitem{ref:cleo}
W.\ Love {\em et al.} (CLEO Collaboration),  Phys.\ Rev.\ Lett. \textbf{101}, 201601 (2008).

\bibitem{ref:bellelfv}
Y.\ Miyazaki {\em et al.} (Belle Collaboration), Phys.\ Lett. \textbf{B660}, 154 (2008).

\bibitem{ref:nussinov}
S.\ Nussinov, R.\ D.\ Peccei and X.\ M.\ Xhang, Phys.\ Rev. D\ \textbf{63}, 016003 (2001).

\bibitem{ref:barbieri}
R.\ Barbieri, L.\ J.\ Hall and A.\ Strumia, Nucl.\ Phys. \textbf{B445}, 219 (1995).

\bibitem{ref:pdg}
C.\ Amsler {\em et al.} (Particle Data Group), Phys.\ Lett. \textbf{B667}, 1 (2008).

\bibitem{ref:thesis}
B.\ Hooberman, Ph.D. Thesis UC Berkeley, SLAC-R-924, (2009).

\bibitem{ref:kk2f}
S.\ Jadach, B.\ F.\ L.\ Ward and Z.\ Was, Comput.\  Phys.\  Commun.\  {\bf  130}, 260 (2000).

\bibitem{ref:bhwide}
S.\ Jadach, W.\ Placzek and B.\ F.\ L.\ Ward, Phys.\ Lett. \textbf{B390}, 298 (1997).

\bibitem{ref:evtgen}
D.\ J.\ Lange, Nucl.\ Instrum.\ Methods {\bf A462}, 152 (2001).

\bibitem{ref:geant}
S.\ Agostinelli {\em et al.} (\textsc{Geant4} Collaboration), Nucl.\ Instrum.\ Methods \textbf{A506}, 250 (2003).

\bibitem{ref:babar}
B.\ Aubert {\em et al.} (\babar\ Collaboration), Nucl.\ Instrum.\ Methods {\bf A479}, 1 (2002).

\bibitem{ref:narsky}
I. Narsky, arXiv:physics/0507143v1 [physics.data-an] (2005).

\bibitem{ref:suppl}
Additional plots are available in the Appendix.

\bibitem{ref:cb}
J.\ E.\ Gaiser, Ph.D. Thesis Stanford U., SLAC-R-255, page 178 (1982).

\bibitem{ref:argus}
H.\ Albrecht {\em et al.} (ARGUS Collaboration), Phys.\ Lett. \textbf{B241}, 278 (1990).

\bibitem{ref:silagadze}
Z.\ K.\ Silagadze, Phys.\ Scripta \textbf{64}, 128 (2001).

\bibitem{ref:black}
D.\ Black {\em et al.}, Phys.\ Rev. D\ \textbf{66}, 053002 (2002).

\end{thebibliography}
\end{document}